\documentclass[11pt]{article}

\usepackage{amssymb}
\usepackage{amsmath}
\usepackage{graphicx}
\renewcommand{\vec}[1]{{\bf #1}}    %%% vectors in bold
\def\beq{\begin{eqnarray}}       %%% begequation/eqnarray
\def\eeq{\end{eqnarray}}        %%% endequation/eqnarray
\def\Box{\square}           %%% Box
\def\diag{\,\mbox{diag}\,}           %%% Box
\def\al{\alpha}
\def\be{\beta}

\def\ga{\gamma}
\def\de{\delta}
\def\ep{\epsilon}

\def\La{\Lambda}
\def\la{\lambda}

\def\si{\sigma}

\def\ph{\varphi}

\def\Ga{\Gamma}

\def\La{\Lambda}

\def\na{\nabla}
\def\pa{\partial}
\newcommand{\rL}{\rho_{\Lambda}}
\newcommand{\CC}{\Lambda}

\newcommand{\rLV}{\rL^{\rm vac}}

\begin{document}

\begin{center}
{\Large\sc On the possible running of the cosmological ``constant''}
 \vskip 6mm

%%%%%%%%%%%%%%%%%%%%%%%%%%%%%%%%%%%%%%%%%%%%%%%%%%%%%%%%%%%%%%%%
\textbf{Ilya L. Shapiro}$^{\,a}$
%%  Also at Tomsk State Pedagogical University, Tomsk, Russia.}
$\,,\,\,$
%%%%%%%%%%%%%%%%%%%%%%%%%%%%%%%%%%%%%%%%%%%%%%%%%%%%%%%%%%%%%%%%
\textbf{Joan Sol\`{a}} $^{\,b}$ \vskip 6mm $^{\,a}$ Departamento de
F\'{\i}sica - Instituto de Ciencias Exatas
\\
Universidade Federal de Juiz de Fora, 36036-330, MG, Brazil \vskip
2mm $^{\,b}\,$
High Energy Physics Group, Dept. ECM and Institut de
Ci{\`e}ncies del Cosmos,\\
    Univ. de Barcelona, Av. Diagonal 647, E-08028 Barcelona, Catalonia, Spain
\vskip5mm

E-mails:  shapiro@fisica.ufjf.br, sola@ecm.ub.es.
\\[0pt]
\end{center}
\vspace{0.0 cm}

\begin{center}
{ABSTRACT}
\end{center}
%\vskip 1mm

\begin{quotation}
\noindent
%% {\bf Abstract.} \
\footnotesize{Despite the many outstanding cosmological observations
leading to a strong evidence for a non-vanishing cosmological
constant (CC) term $\CC$ in the gravitational field equations, the
theoretical status of this quantity seems to be lagging well behind
the observational successes. It thus seems timely to revisit some
fundamental aspects of the CC term in Quantum Field Theory (QFT). We
emphasize that, in curved space-time, nothing a priori prevents this
term from potentially having a mild running behavior associated to
quantum effects. Remarkably, this could be the very origin of the
dynamical nature of the Dark Energy, in contrast to many other
popular options considered in the literature. In discussing this
possibility, we also address some recent criticisms concerning the
possibility of such running. Our conclusion is that, while there is
no comprehensive proof of the CC running, there is no proof of the
non-running either. The problem can be solved only through a deeper
understanding of the vacuum
contributions of massive quantum fields on a curved space-time
background. We suggest that such investigations are at the heart of
one of the most important endeavors of fundamental theoretical
cosmology in the years to come.}
\\
%{\bf Keywords:} Quantum field theory in curved spacetime,
%Cosmological constant, Renormalization group.
%\\
%{\bf PACS:} $\,$
%04.62.+v %%% Quantum field theory in curved spacetime
%,$\,\,$
%11.10.Gh %%% Renormalization
%,$\,\,$
%11.10.Hi %%% Renormalization group evolution of parameters
\end{quotation}

%% \newpage
%%%%%%%%%%%%%%%%%%%%%%%%%%%%%%%%%%%%%%%%%%%%%%%%%%%%%%%%%%%%

\section{\large\bf Introduction}

The relevance of the CC problems\,\cite{weinberg89,IRGAC06} has
triggered a renewed interest on the dynamical quantum effects on the
vacuum energy density and their possible implications in cosmology.
The standard way to parameterize the leading quantum effects is the
renormalization group (RG). The theoretical background for the RG
running of the CC has been established in the papers
\cite{{NelPan},{buch84},{Toms83}} where this running was considered
in the framework of Quantum Field Theory (QFT) in curved space-time,
and also in \cite{frts82} where the underlying theory was Quantum
Gravity (QG). From the viewpoint of the physical interpretation of
the RG, one can identify three approaches dealing with the CC
problem: one of them can be described as an attempt to solve the
``old'' CC problem\,\cite{weinberg89} via the RG screening at low
energies \cite{Pol81,TV,antmot,lam,jackiw}. Another is the
functional renormalization group approach to
QG\,\cite{Reuter1,Reuter2}, which is based on the Wilsonian notion
of average action and its RG equation; it is applied in the
framework of QG and is formulated non-perturbatively\,\cite{RW}.
Finally, there is the approach proposed by the present authors
\,\cite{cosm,nova}, in which it is explored the possibility of
having a relatively moderate running of the CC within perturbative
QFT in a curved background\,\cite{IRGAC06}. The latter kind of
approach is closer to the methods of Particle Physics phenomenology.
In this same line, we have the
contributions\,\cite{babic,CCfit,Gruni,JSHSPL05,JSS1,fossil,FBauer1}.
In all these cases, one studies the running of the CC without making
direct reference to the old CC problem \cite{weinberg89}. Let us
also point out that most of the aforesaid RG papers use the Minimal
Subtraction (MS) scheme of renormalization, which is a pretty well
established albeit not directly physical procedure, whereas a few
recent works have dealt with the yet undeveloped (despite badly
needed) physical scheme of renormalization in cosmology. This
scheme, when a rigorous formulation becomes finally available,
should appropriately extend the existing physical renormalization
schemes \cite{Manohar96} into curved space-time.

In the papers \cite{cosm,nova}, we suggested the possibility of the
CC running, although this cannot be rigorously proven in the context
of the present day QFT in curved space. Recently, an attempt to
prove that such running is mathematically impossible was undertaken
in\,\cite{AG}. However, the argumentation presented in that work is
misleading. This conclusion will ensue here as a corollary of our
general discussion (see also \,\cite{DoesCCrun} for more
details)\,\footnote{See also the recent references
\cite{Bauer2,BFLWard}, which also consider that the arguments of
\,\cite{AG} cannot be supported.}. Our main aim in this Letter is of
general nature; we wish to critically assess the physical conditions
leading to the CC running in QFT, and also to emphasize that this
issue could be of paramount importance for the theoretical
cosmology. This is especially so after the many efforts devoted
during the last quarter of a century trying to unsuccessfully
replace Einstein's CC with a variety of ersatz entities of different
nature without, unfortunately, improving significantly the overall
status of the difficult CC problem(s)\,\cite{weinberg89}.

The paper is organized as follows. In section 2 we briefly
discuss the CC and the CC problem. In particular, we
demonstrate the importance of the curved-space metric and
remember that in the conformal parameterization the Einstein-Hilbert
action with CC is equivalent to the scalar field action
and that the renormalization of the CC term is analogous
to the one of the scalar potential term in the Nambu-Jona-Lasinio
(NJL) model. In section 3 we briefly review the notion of
running in QFT. Section 4 is devoted to the detailed analysis
of the criticism of our previous papers in \cite{AG}. In section
5 we discuss the unusual possible form of the effective action
of the metric which can be responsible for the running and
show that, despite very strong restrictions, such running can
not be ruled out. Finally, in section 6 we draw our conclusions.

%% 3
%%%%%%%%%%%%%%%%%%%%%%%%%%%%%%%%%%%%%%%%%%%%%%%%%%%%%%%%%%%%
%%%%%%%%%%%%%%%%%%%%%%%%%%%%%%%%%%%%%%%%%%%%%%%%%%%%%%%%%%%%
%\section{\large Background notions}

%% 3.1
%%%%%%%%%%%%%%%%%%%%%%%%%%%%%%%%%%%%%%%%%%%%%%%%%%%%%%%%%%%%
\section{\large What is $\CC$, and what are the $\CC$ problems}

Before starting to discuss the CC running, it is worthwhile to
remember what is the CC term $\CC$ and what is the RG running. The
modern gravitational physics starts from the action of the form
\beq S_{total} = -\frac{1}{16\pi G}\,\int d^4 x\,\sqrt{-g} \,\left(
R + 2\La \right) + S_{HD} + S_{matter}\,. \label{total} \eeq
Here the first term is the Einstein-Hilbert action with the
cosmological constant $\La$; the second term $ S_{HD}$ includes
higher derivatives (cf. section 5), which are necessary for the
consistency of a quantum theory in curved space (see, e.g.,
\cite{birdav}, \cite{book} or \cite{PoImpo} for an introduction) and
the last term, $S_{matter}$, represents the action of matter,
responsible for the energy-momentum tensor $T_\mu^\nu$.  In the low-energy domain, one can in principle
disregard $S_{HD}$ and the dynamical equations for the metric take
on the Einstein form
\beq R_\mu^\nu - \frac12\,R\,\de_\mu^\nu \,=\,8\pi G\,T_\mu^\nu +
\La\,\de_\mu^\nu\,. \label{Einstein} \eeq
If we consider matter as an isotropic fluid with energy density
$\rho$ and pressure $p$, the energy-momentum tensor has the
following form (in the locally co-moving frame):
 \beq T^\nu_\mu=\diag
\left(\rho,\,-p,\,-p,\,-p\right)\,. \label{EMT} \eeq
It is easy to see that the $\La$-dependent term in (\ref{Einstein})
has exactly the form (\ref{EMT}), with the ``vacuum energy density''
$\rho_\Lambda^{vac}$ and ``vacuum pressure'' $p_\Lambda^{vac}$ being
\beq \rho_\Lambda^{vac} = \frac{\Lambda}{8\pi G} = -
p_\Lambda^{vac}\,. \label{CCd2} \eeq
Thus, the vacuum part of the energy-momentum tensor is
${\left(T^{vac}\right)}^\nu_\mu=
\rho_\Lambda^{vac}\,\delta^\nu_\mu$, which justifies to call
(\ref{CCd2}) the ``vacuum energy''.

Let us start the discussion of the CC term by making an important
remark. Without gravity the CC term is nothing but an irrelevant
constant. The CC acquires dynamical significance only through the
Einstein equations (\ref{Einstein}), which tell us the space-time is
curved. To better understand the role of the curved space for the CC
term, let us consider another parametrization of the metric
\beq g_{\mu\nu} = \frac{\chi^2}{M_P^2}\, {\bar g}_{\mu\nu}\,,
\label{conf_rep} \eeq
where ${\bar g}_{\mu\nu}$ is some fiducial metric with fixed nonzero
determinant; for instance, it can be the flat metric $\,{\bar
g}_{\mu\nu}=\eta_{\mu\nu}$. Furthermore, $\chi=\chi(x)$ is a scalar
field which can be identified as a conformal factor of the metric,
and $M_P\equiv G^{-1/2}$ is the Planck mass. It is easy to see that
the CC term looks rather different in these new variables:
\beq S_\Lambda \,=\, -\, \int d^4x \sqrt{- g}\, \frac{\Lambda}{8\pi
G} \,=\, -\, \int d^4x \sqrt{-{\bar g}}\,f \chi^4\,, \qquad
\mbox{where} \qquad f = \frac{\Lambda}{8\pi G\,M_P^4} =
\frac{\La}{8\pi\,M_P^2}\,. \label{CCd1} \eeq
The last expression is nothing but the usual quartic term in the
potential for the scalar interaction. One may note that, under the
same change of variables, the Einstein-Hilbert term transforms into
the action of the scalar field $\chi$ with the negative kinetic term
and nonminimal conformal coupling to curvature
\cite{deser,sola8990,conf}. Furthermore, the massive term in the
spinor Lagrangian becomes a Yukawa-type interaction between the
fermion and the scalar degree of freedom of the metric $\chi$.

One important consequence which follows from the above observation
is that the renormalization of the Newton constant and the CC due to
the quantum effects of the spinor field is completely similar to the
renormalization in the well known NJL model. In
this model, one does not quantize the scalar field, exactly as we do
not intend to quantize the metric in QFT in curved space-time.
Furthermore, within the MS scheme of renormalization, the RG running
of the CC in the theory of spinor field is mathematically equivalent
to the one of the parameter $\,f\,$ in the NJL model \cite{NJL}.
Similar equivalence can be easily established for the quantum
effects of a free massive scalar field $\ph$ -- in this case, the
quantum field interacts with $\,\chi\,$ via the $\ph^2\chi^2$-term.

On the phenomenological side, the CC is the most natural candidate
for the Dark Energy (DE) which is responsible for the cosmic
acceleration. However, all models of the DE must face the so-called
``old CC problem''\cite{weinberg89}, i.e. the formidable task of
trying to understand the enormous ratio $r=\rho_{\Lambda}^{\rm
QFT}/\rho_{\Lambda}^0$ between the theoretical QFT computation of
the vacuum energy density and its presently observed value,
$\rho_{\Lambda}^0\sim 10^{-47}\,\text{GeV}^4$, obtained from modern
cosmological data\,\cite{cosmdata}. For instance, the Standard Model
contribution from electroweak interactions generates the huge ratio
$r\sim 10^{55}$\,\cite{nova}. If that is not enough, there is
another pressing CC problem, the ``coincidence problem''; namely, to
understand why $\rho_{\Lambda}^0$ is precisely of the same order of
magnitude as the current matter density $\rho_m^0$. Observations
indeed provide
$\rho_{\Lambda}^0/\rho_m^0=\Omega_{\Lambda}^0/\Omega_{m}^0\simeq
7/3={\cal O}(1)$. This is of course very puzzling because $\rho_m$
decays fast as $\rho_m\sim 1/a^3$ at any epoch. So, why on earth is
$\rho_m$ almost equal to $\rho_{\Lambda}$ at our living epoch (i.e.
at $a=1$)?

It is generally accepted that the coincidence problem could be
ameliorated if the DE would be a dynamical quantity, e.g. related to
some scalar field. Alternatively, it could be that the CC term is
not really constant and varies together with some cosmic parameter.
At the moment, only phenomenological models are available on all
these options\, \cite{Quintessence,Overduin}. Most definitely, a
more fundamental possibility would be that the CC were a dynamical
variable tied to genuine quantum effects. In such case, the RG
should be the most appropriate theoretical instrument to explore
this framework \cite{Pol81}-\cite{Reuter2}\,\cite{cosm,nova} and
elucidate its phenomenological
consequences\,\cite{babic}-\cite{fossil}, including its possible
connection with the coincidence problem\,\cite{DEpert}.

A fundamental equation playing an important role here is the
covariant energy conservation. In the Friedmann-Lema\^\i
tre-Robertson-Walker (FLRW) cosmological context, the Bianchi
identity fulfilled by the Einstein's tensor on the \textit{l.h.s.}
of Eq.\,(\ref{Einstein}) implies the following relation among the
various terms on its \textit{r.h.s.} :
\beq \label{BianchiGeneral}
\frac{d}{dt}\,\big[G(\rho_\Lambda+\rho_m)\big]+3\,G\,H\,(\rho_m+p_m)=0\,,
\eeq
where $H=\dot{a}/a$ is the expansion rate or Hubble function. This
equation does not rule out neither $H$-dependence nor $a$-dependence
of the CC, which can take place either because of the energy
exchange between vacuum and matter or due to the variable Newton
``constant'' $G$ \cite{Gruni,JSHSPL05,fossil}. In both cases, there
is the possibility for a time-evolving cosmological term,
$\dot{\rho}_\Lambda\neq 0$, which can be a natural alternative to
the dynamical DE models exclusively based on ad hoc scalar
fields\,\cite{Quintessence}.

The above consideration shows that the way leading to CC running is,
in principle, open and perfectly consistent with covariance.
However, are the quantum effects ultimately responsible for the CC
dynamics? At present, our knowledge of QFT in curved space-time is
not sufficient to give a definite answer to this question
\cite{nova,PoImpo}. There is, however, a strong hint in favor of the
existence (despite the complicated form) of the quantum corrections
under discussion. All kinds of such corrections can be seen as
contributions to the effective action of gravity, where the CC plays
a central role at low energies. Therefore, they can be considered in
the framework of the induced gravity paradigm \cite{Adler}. It is
well known that the induced gravitational action {\it always} has a
one-parameter ambiguity \cite{Adler,book}. The parameter behind this
ambiguity must have some physical sense and, in the cosmological
setting, it may be related to $a(t)$, $H(t)$ or some combination.
This relation opens the door to the physically relevant RG running
of the various terms in the vacuum action, including the CC.

%% 3.2
%%%%%%%%%%%%%%%%%%%%%%%%%%%%%%%%%%%%%%%%%%%%%%%%%%%%%%%%%%%%
\section{\large What means RG running in QFT and in cosmology}

Conventionally, the classical theory starts from establishing its
action $S$. The quantum theory, instead, is often characterized by
the $S$-matrix elements. However, in the QFT framework, it is also
advantageous to use the effective action (EA) $\Gamma$ of the mean
fields (see, e.g., \cite{Coleman85,Brown}), which can be looked upon
as a generalization of the classical action in the quantum domain.
This approach becomes especially significant in the presence of
gravity, where the definition of the $S$-matrix can be problematic.
Taking variations of the EA, one arrives at the Green functions and
finally to the amplitudes. At this point it is important to remember
that, in contrast to the classical action, the EA always has certain
ambiguities, which eventually disappear in the amplitudes. These
ambiguities have a manifold origin: they may come e.g. from the
choice of the parametrization (or gauge fixing, as a particular
case) of the quantum fields; or from the choice of the
renormalization scheme; and also from the dependence of the various
parts of the EA on the renormalization parameter $\mu$. Indeed, the
latter can be a rather artificial quantity, as e.g. in the case of
the MS scheme, but in some cases it can be chosen more physically
and represent e.g. an arbitrary subtraction point in momentum
space\,\cite{Coleman85}.

The derivation of the EA and working out its ambiguities can be
regarded as the main target of QFT. In general, one can not
completely calculate the EA, which is typically given by a non-local
and non-polynomial expression in the mean fields. Quite often there
is a special sector of the EA (viz. the one containing the leading
quantum effects) which can be easily accounted for because these
effects become just parameterized by the aforementioned arbitrary
mass scale $\mu$. The appearance of this scale is characteristic of
the renormalization procedure in QFT owing to the intrinsic breaking
of scale invariance by the quantum effects. The $\mu$-dependence
shows up in many places in the EA, and in particular also in certain
non-local parts related to divergences. Despite the neat cancelation
of the overall $\mu$-dependence in the EA, the different quantum
parts are parameterized by $\mu$, and this apparently innocent fact
is absolutely crucial as it enables us to restore the structure of
form factors, namely the pieces that transport the physical
information of the quantum effects. E.g., in the one-loop QED with
zero masses (or, equivalently, at very high energy), the
electromagnetic part of the EA has the form
\beq \Ga^{(1)}_{em}\,=\,-\frac{1}{4e^2(\mu)}\, \int
d^4x\,F_{\mu\nu}\,\Big[\,1 -  \frac{e^2(\mu)}{12\pi}\,\ln
\Big(-\frac{\Box}{\mu^2}\Big) \Big]\,F^{\mu\nu}\,, \label{QED} \eeq
where $e(\mu)$ is the renormalized QED charge in the MS scheme.  It
is apparent from (\ref{QED}) that the effective (or ``running'') QED
charge in momentum space satisfies
\begin{equation}\label{renormchargeMSQEDHE}
\frac{1}{e^2(Q^2)}=\frac{1}{e^2(\mu^2)} - \frac{1}{12\pi^2}\,
\ln\left(\frac{|Q^2|}{\mu^2}\right)\,,\ \ \ \ \ \ (|Q^2|\gg
m_e^2)\,.
\end{equation}
This relation also follows upon integrating the differential
equation
\begin{equation}\label{RGEQED}
\mu\frac{d\,e(\mu)}{d\mu}=\frac{e^3}{12\,\pi^2}\equiv
\beta_{e}^{(1)}
\end{equation}
from $\mu$ to ${\mu'}$ and then replacing ${\mu'}^2\to |Q^2|$.
Here $\beta_{e}^{(1)}$ is the $\beta$-function of QED at
one-loop in the MS. Equation (\ref{RGEQED}) is correspondingly
called the RG equation of the renormalized charge in the MS
scheme.  Notice from (\ref{renormchargeMSQEDHE}) that $e(Q^2)$
increases with $|Q^2|$, as expected from the non-asymptotically
free character of QED -- a well tested feature of this theory.

The remarkable property of the $\mu$-dependence is that the
high-energy limit of the theory is reproduced, in the leading
approximation, by taking $\mu \to \infty$. Due to the simplicity of
the form factor for the massless case (the above expressions are a
nice illustration) one can always restore such form factor using the
$\mu$-dependence. In other words, using this ``RG-trick'' one can
immediately retrieve the non-trivial structure of
the form factor as a function of the $\Box$-operator. In momentum
space, the (artificial) $\mu$-dependence paves our way to discover
the explicit (and physical) $Q$-dependence
(\ref{renormchargeMSQEDHE}). In fact, this is the essential and
practical aspect of the RG-method in Particle Physics: being the
combination $Q^2/\mu^2$ the natural variable in the renormalized
scattering amplitudes, the RG helps us to find out physical quantum
effects that ``run'' with the energy $Q$ (or some external field) by
just inspecting the $\mu$-parameterization inherent to the various
parts of the S-matrix and EA.

The simple QED example above can be generalized for any parameter in
QFT, as is well-known from standard RG arguments\,\cite{Coleman85}.
In particular, it also applies to the CC in flat space as far as the
$\mu$-dependence is concerned\,\cite{Brown}. However, whereas the
transition from (\ref{RGEQED}) to the physical $Q$-running
(\ref{renormchargeMSQEDHE}) is well-established in Particle Physics,
the situation with the CC is more delicate. To start with, if the RG method
can be applied in cosmology, it should mean that the vacuum energy
density becomes a running parameter. Therefore, as in the previous
case (\ref{RGEQED}), there should exist a fundamental RG equation of
the form
\begin{equation}\label{RGCC}
\mu\frac{d\rL}{d\mu}=\beta_{\La}(P, \mu)\,,
\end{equation}
which is supposed to describe the leading quantum contributions to
it, where $\beta_{\La}$ is a function of the parameters $P$ of the
effective action (EA).  The quantity $\rL$ in (\ref{RGCC}) is a
($\mu$-dependent) renormalized part of the complete QFT structure of
the vacuum energy. While the full vacuum energy is a physical
observable, and hence overall $\mu$-independent, those parts of it
related to quantum effects are still parameterized by $\mu$ and
hence may contain relevant information on the physical running of
the form factors (similar to the QED example mentioned above),
except that here the correspondence of $\mu$ with a physical
quantity ($Q$ in the QED case) is not transparent in the
cosmological setting. We have said that the RG technique can
actually be extended to the whole Particle Physics domain; yet, not
even in this case can one arbitrarily choose a subtraction scheme
for a particular calculation. For instance, in QCD one cannot choose
the on-shell scheme (quarks are never free particles on the mass
shell), and so here we must content ourselves with unphysical
off-shell schemes, such as the MS. But this is no obstacle for
computing physically meaningful observables at the level of
amplitudes and cross-sections, basically because the UV recipe
$\mu\to Q$ is absolutely unambiguous and robust.

In cosmology the situation is more complicated, and the final
contact with physics is correspondingly more subtle. The root of the
difficulty is partly because (as remarked above) the physical scale
behind the quantum effects is not obvious. Moreover, when addressing
the cosmology of our present Universe, we are actually dealing with
the infrared domain, and hence one is unavoidably led to consider
the massive case and the decoupling effects. Obviously this cannot
be automatically handled in the MS scheme, as the MS violates the
decoupling theorem\,\cite{AC}. The simplest option is to proceed as
in QCD, namely to assume a ``sharp cut-off'', which means to
disregard completely the contributions of massive fields at the
energy scale below their mass and, at the same time, treat their
contributions above the proper mass scale as high-energy ones,
without taking decoupling into account. This was exactly the option
adopted in\,\cite{cosm}.

A crucial question is: what is the relevant physical quantity
playing the role of $Q$ in cosmology? The answer is neither obvious
nor unique, as there are several reasonable candidates. Yet, it
must be linked to the existence of a nontrivial external metric
background\,\cite{Reuter1,Reuter2,cosm,nova}. For instance, the
dynamical properties of this curved background (viz. the expanding
FLRW space-time) can be characterized by the expansion rate $H$. If
so, they are expected to induce a functional dependence $\rL=\rL(H)$
in which, as usual, the quantum effects should be parameterized by
some renormalization scale $\mu$. Lacking, however, at present of
the suitable techniques to tackle a full-fledged computation of the
decoupling effects, one can at least take the general covariance as
the main guide. Overall, it suggests that the solution of the RG
equation (\ref{RGCC}) should lead to an even-power law of the Hubble
expansion rate:
\begin{equation}
\label{runphysCC} \rL(H)= \rL^{0}+ \frac{3\,\nu}{8\pi}\,M_P^2\,
\left(H^2-H^2_0\right)+ {\cal O}(H^4)\,,
\end{equation}
where $\rL^{0}$ and $H_0$ are the current values of the CC and the
Hubble rate respectively, and $\nu$ is a coefficient playing the
role of $\beta$-function. It is of course understood that the ${\cal
O}(H^4)$ effects are negligible in the current Universe. This
expression is precisely of the ``soft-decoupling'' form first
introduced in \cite{nova}. The above expression is the kind of
running law that has been tested and further elaborated in many
subsequent papers on the CC running
\cite{babic,CCfit,Gruni,JSHSPL05,JSS1,fossil,FBauer1,BPS09a}.

Let us summarize the general features which are indispensable for a
proper understanding and correct use of the RG. Even though this is
standard, it is nevertheless necessary, since it is a source of
considerable confusion in the recent literature, specially when
applied to cosmology\,\cite{AG}. The basic doctrine of the RG is
that the full renormalized effective action, $\Ga$, whether in the
pure realm of QFT, or in the context of cosmology, or for that
matter in any other conceivable context of theoretical physics, must
not depend on the numerical value of the arbitrary renormalization
scale $\mu$. Moreover, this is true in the MS scheme or in any other
subtraction scheme (e.g. the momentum subtraction scheme, which is
more physical and closer to the on-shell scheme). This property
belongs to the very fundamental definition of the renormalized EA,
which must always be equal to the bare EA. Let us e.g. assume
dimensional regularization (with $n$ space-time dimensions) and let
$\Phi$ and $P$ be the full sets of fields and parameters,
respectively. If we denote the bare quantities with a subindex $0$,
we have
\beq \Ga_0[g_{\al\be},\Phi_0,P_0,n]
\,=\,\Ga[g_{\al\be},\Phi,P,n,\mu] \ \ \ \Longleftrightarrow\ \ \
\mu\,\frac{d\,\Gamma}{d\mu}=0\,. \label{nn8} \eeq
This relation holds in both flat and curved space time and for all
kinds of renormalizable theories. Here the $\mu$-dependence appears
in order to restore the canonical dimensions of the renormalized
fields and parameters on the \textit{r.h.s.} of (\ref{nn8}), and
manifests itself in the following two ways:

1) All renormalized fields $\Phi$ and parameters $P$ depend on
$\mu$: in Eq.(\ref{nn8}) it is implicitly understood that
$\Phi=\Phi(\mu)$ and $P=P(\mu)$ ;

2) The functional form of EA depends explicitly on $\mu$.

These two $\mu$-dependencies always cancel perfectly, for otherwise
the renormalized EA would be overall $\mu$-dependent, violating its
own definition (\ref{nn8}). So, if we take together the two types of
$\mu$-dependencies, we just eliminate entirely the possibility to
use the RG as a tool to explore the structure of the quantum
corrections. The use of the RG requires, therefore, that the
implicit and explicit $\mu$-dependencies 1) and 2) are used
separately.

In particular, in the MS, being $\mu$ an artificial mass unit, its
practical use always requires an additional effort. In the flat
space theory we meet two different standard interpretations of
$\mu$, each of them implying some relation between MS and other,
more physical, renormalization schemes in the corresponding limit.
In the case 1) above we have to look for the correspondence with the
momentum subtraction scheme at high energies. The case 2), instead,
is mainly used for the analysis of phase transitions, meaning that
we associate $\mu$ with the mean value of the almost static (usually
scalar) field and arrive at the interpretation of the
$\mu$-dependent effective potential (see next section).

The use of $\mu$ is justified only if we are clearly understanding
which physical parameter is behind it. For instance, in the case of
(\ref{QED}) we associate  $\mu$ with the value of a typical momentum
$Q$ in the scattering amplitudes and arrive at the $Q$-dependence of
the coupling constant, see Eq.(\ref{renormchargeMSQEDHE}). Similar
dependence leads to the asymptotic freedom in the non-Abelian gauge
theories. Obviously, the correspondence of $\mu$ with $Q$ (or any
other physical parameter) can be performed {\it only if} the $\mu$
dependence is retained in the running parameters, but it would be
impossible if one would eliminate $\mu$ by canceling it with the
other (explicit) $\mu$-dependences associated to the functional
dependence of the EA!

At present, the use of the MS-based RG for the CC and, in general,
for the gravitational applications, is fairly nontrivial. The reason
is simple but categorical: in the presence of a non-vanishing CC,
the space-time can not be flat, and therefore the calculations
cannot be performed by expanding around a flat background (which is
the only technique we know to perform calculations in
practice\,\cite{apco}). In this situation, the $\mu$-dependence can
be hardly associated to some scattering amplitudes because the
S-matrix cannot be straightforwardly defined for non-flat spaces.
Furthermore, except for the simplest static spaces, the definition
of the effective potential is not straightforward in curved spaces
\cite{sponta}. Finally, in curved space the almost static scalar
field does not mean, in general, to have static metric or static
curvature.

%% 4
%%%%%%%%%%%%%%%%%%%%%%%%%%%%%%%%%%%%%%%%%%%%%%%%%%%%%%%%%%%%
\section{\large Effective potential and running $\CC$}
%% : critical analysis of reference \cite{AG}}

The effective potential\,\cite{Coleman85}, being the quantum analog
of the classical potential, appears as the natural tool to explore
the vacuum energy in QFT, and one may suspect that its properties
have an important bearing on the CC problem. However, to understand
the precise connection between the two concepts is not so simple as
one might naively think. Therefore, it seems necessary to clarify
this point in order to avoid misleading conclusions\,\cite{AG}. Let
us consider the RG equation for the effective potential $V$ of the
real massive scalar field with the $\la\phi^4$ interaction. The CC
can be included into the scalar potential in the form of an additive
$hm^4$ term, where $m$ is the scalar mass and $h$ is an independent
dimensionless parameter. The RG equation for $V$ follows from the
fact that, for constant mean fields and in the local approximation,
the effective action boils down to $\Gamma=-V\,\Omega$, where
$\Omega$ is the space-time volume. The RG-invariance of $\Gamma$,
expressed by Eq. (\ref{nn8}), leads to the following PDE for the
renormalized effective potential:
\beq \Big(\mu\frac{\pa}{\pa \mu}+\be\frac{\pa}{\pa \la} +\ga_m m^2
\frac{\pa}{\pa m^2} +\ga_\phi \phi \frac{\pa}{\pa \phi} + \be_h
\frac{\pa}{\pa h}\Big)V(\phi,m^2,\la,h,\mu)\,=\,0\,. \label{RG} \eeq
It would be incorrect to think of this equation as being tantamount
to a kind of non-running theorem for the vacuum energy and the CC.
This way of thinking could only be the result of a misleading
interpretation\,\cite{AG} about the significance of the RG and of
the notion of vacuum energy in the presence of gravity. Indeed,
neither the effective potential nor its vacuum expectation value can
be identified with the vacuum energy in the cosmological context
(cf. section 5). Put another way: in flat space, $V$ cannot provide
by itself any crucial insight on the possible running of the CC.
Yet, this does not mean that we cannot extract valuable information
using the RG technique if we, first of all, embed this theory in a
non-trivial external background\,\cite{cosm,nova}. Let us, thus,
analyze the situation more carefully, step by step:
%%%%%%%%%%%%%%%%%%%%%%%%%%%%%%%%%%%%%%%%%%%%%%%%%%%%%%%%%%%%%
%%%%%%%%%%%%%%%%%%%%%%%%%%%%%%%%%%%%%%%%%%%%%%%%%%%%%%%%%%%%%
\vskip 1mm

 {\Large $\bullet$} \ First of all a technical point:
Eq.\,(\ref{RG}) is, in fact, a sum of the two independent equations,
to wit, one can split the overall effective potential as a sum of
two pieces, the $\phi$-independent (vacuum) term and the
$\phi$-dependent (scalar) term, as follows:
\beq V(\phi,m^2,\la,h,\mu)=V_{scal}(\phi,m^2,\la,\mu) +
V_{vac}(m^2,\la,h,\mu) \label{Veff}\,. \eeq
In this way, we can also split the RG equation (\ref{RG}) into two
independent RG identities:
\begin{eqnarray}
&&\Big(\mu\frac{\pa}{\pa \mu} + \be_{\la}\frac{\pa}{\pa \la}
+\ga_m m^2 \frac{\pa}{\pa m^2} +\ga_\phi \phi \frac{\pa}{\pa \phi}
\Big)V_{scal}(\phi,m^2,\la,\mu)\,=\,0\,,
\\
&&\Big(\mu\frac{\pa}{\pa \mu} + \be_{\la} \frac{\pa}{\pa \la}+\ga_m
m^2 \frac{\pa}{\pa m^2}+\be_h \frac{\pa}{\pa
h}\Big)V_{vac}(m^2,\la,h,\mu) \,=\,0\,.
\label{RG-sep}
\end{eqnarray}
In order to understand the origin of this splitting, one has to
introduce the functional called effective action of vacuum
$\Ga_{vac}$. It is that part of the overall EA which is left when
the mean scalar field $\phi$ is set to zero:
$\Ga_{vac}=\Ga[\phi=0]$. Thus, it is a pure quantum object
which only depends on the set of parameters $P=m,\lambda,...$
of the classical theory. At the functional level,
\beq e^{i\Ga_{vac}}\,=\,\int {\cal D}\phi \ e^{iS[\phi; J=0]}\,,
\label{vacEA2} \eeq
where the source $J$ is set to zero. In this way, the functional
$\Ga_{vac}$ is the generator of the proper vacuum-to-vacuum
diagrams. In flat space, these diagrams are removed by normalizing
the functional to one. But in curved space they are significant.
From the RG-invariance of the renormalized EA -- see
Eq.\,(\ref{nn8}) -- it follows immediately the $\mu$-independence of
the renormalized functional $\Ga_{vac}$ and, therefore, we arrive at
the second identity (\ref{RG-sep}) for the vacuum part of the
effective potential, while the first identity is the result of the
subtraction of (\ref{RG-sep}) from (\ref{RG}). The net result is
that the vacuum and matter parts of the effective potential are
overall $\mu$-independent separately and no cancelation between them
can be expected. \vskip 1mm

{\Large $\bullet$} \ The $\mu$-independence of the EA holds by
construction and definitely does not mean that there is no physical
running, as it was assumed in \cite{AG}. As we have already discussed
in section 3, and as reflected by Eq, (\ref{RG}), the explicit
$\mu$-dependence of the function $V(\phi,m^2,\la,h,\mu)$ is
automatically canceled by the $\mu$-dependences of the field $\phi$
and parameters $\la,\,m,\,h$. Using both dependencies at the same
time makes no sense, just because we know in advance that they
cancel. This is true for {\it any} renormalizable theory and for
{\it any} sector of such theory, in particular for the NJL model,
QED and QCD. Recall e.g. that for the NJL model the renormalization
can be simply linked with the CC case just by changing the
parametrization of the external metric, see Eq.\,(\ref{CCd1}).
%%%%%%%%%%%%%%%%%%%%%%%%%%%%%%%%%%%%%%%%%%%%%%%%%%%%%%%%%%%%%
%%%%%%%%%%%%%%%%%%%%%%%%%%%%%%%%%%%%%%%%%%%%%%%%%%%%%%%%%%%%%
\vskip 1mm

 {\Large $\bullet$} \ In the vacuum sector, the one-loop effects
modify the relation (\ref{CCd2}) as follows:
\begin{equation}
\label{Vacfree} V_{vac}=\rLV(\mu)+\delta\rLV+{\bar V}_{vac}^{(1)}\,.
\end{equation}
For a scalar field with mass $m$, the dimensionally regularized
form of the vacuum-to-vacuum diagram in flat space gives
\begin{equation}\label{Vacfree2}
{\bar V}_{vac}^{(1)}
 = \frac12\,\mu^{4-n}\, \int\frac{d^{n-1} k}{(2\pi)^{n-1}}
   \,\sqrt{\vec{k}^2+m^2}
= \frac12\,\be_\La^{(1)}\,\left(-\frac{2}{4-n}
-\ln\frac{4\pi\mu^2}{m^2}+\gamma_E-\frac32\right) \,,
\end{equation}
where $n\rightarrow 4$ in the last expression. The one-loop
$\beta$-function of the vacuum $\CC$ term reads\,\cite{nova}
\begin{equation}
\label{beta4}
\be_\La^{(1)}=\frac{m^4}{32\,\pi^2}\,.
\end{equation}
Equation (\ref{Vacfree2}) is divergent and needs a subtraction. If
we adopt the $\overline{MS}$ subtraction scheme, the counterterm
$\delta\rLV$ gets fixed in such a way that the renormalized result
is
\begin{equation}
\label{rfvacenergy2}
V_{vac}\,=\,\rLV(\mu)+ {\bar V}^{(1)}_{vac}
\,=\,\rLV(\mu)+\frac{1}{2}\,\be_\La^{(1)}
\,\left(\ln\frac{m^2}{\mu^2}-\frac32\right)\,.
\end{equation}
Plugging the $\beta$-function (\ref{beta4}) on the \textit{r.h.s} of
the RG equation (\ref{RGCC}), we can easily integrate it (because at
one-loop the mass $m$ does not run with $\mu$):
\begin{equation}\label{intrLV}
\rLV(\mu)=\rho^{\rm
vac}_{\La}(\mu_0)-\frac{1}{2}\,\be_\La^{(1)}\,\ln\frac{\mu_0^2}{\mu^2}\,.
\end{equation}
One can check that, after replacing Eq.\,(\ref{intrLV}) into the
Eq.\,(\ref{rfvacenergy2}), we arrive at the expression in which
$\mu$ is replaced by $\mu_0$. This is the realization, in this
particular example, of the RG-invariance of the vacuum part of the
effective potential, i.e. of Eq.\, (\ref{RG-sep}). The procedure can
be easily extended to any loop order and $V_{vac}(m,\lambda)$ is in
general a function of both the mass $m$ and of the self-coupling
$\lambda$.\vskip 2mm
%%%%%%%%%%%%%%%%%%%%%%%%%%%%%%%%%%%%%%%%%%%%%%%%%%%%%%%%%%%%%
%%%%%%%%%%%%%%%%%%%%%%%%%%%%%%%%%%%%%%%%%%%%%%%%%%%%%%%%%%%%%
\vskip 1mm

 {\Large $\bullet$} \ The situation for the induced vacuum energy
density is similar. The form of the effective potential $V$ for the
scalar field with the classical potential $U(\phi)$ is well-known
(see, e.g., \cite{Coleman85}):
\beq V(\phi)=U(\phi)+ {\bar V}^{(1)}(\phi) \,=\,U(\phi) +
\frac{1}{64\pi^2}\,{U^{\prime\prime}}^2 \,\Big(\ln
\frac{{U^{\prime\prime}}^2}{\mu_0^2} -\frac12\Big)\,. \label{pot}
\eeq
In the case of
\beq U(\phi) = \frac{m^2 \phi^2}{2}+\frac{\la
\phi^4}{4!}\,, \label{clpo} \eeq
the one-loop correction yields
\beq {\bar V}^{(1)}(\phi)\,=\, \frac{1}{64\pi^2}\,\Big(m^2+\frac{\la
\phi^2}{2}\Big)^2 \,\left[\ln \frac{ \big(m^2+{\la
\phi^2}/{2}\big)}{\mu^2} -\frac32\right]\,. \label{effpo1} \eeq
To obtain $V_{scal}(\phi,m^2,\la,\mu)$, we just subtract from
(\ref{pot}) the vacuum part at one-loop, which is given by the
second term on the \textit{r.h.s.} of (\ref{rfvacenergy2}) -- and
also by the value of $V(\phi)$ at $\phi=0$,
Eq.\,(\ref{rfvacenergy2}). The $\phi$-dependent part of the
potential $V_{scal}(\phi,m^2,\la,\mu)$ is just given by the
difference $V(\phi)-V(\phi=0)$. The fact that it depends explicitly
on $\mu$ and, at the same time, satisfies (17) is because there is
still the implicit $\mu$-dependence that is associated to the
renormalization of the parameters $\la$ and $m$. Once the two types
of $\mu$-dependences meet together, they annihilate each other.

In order to give one more illustrative example, let us concentrate
on the massless case. The RG equation for $\la$ has a form similar
to (\ref{RGEQED}) in the massless QED case:
\beq
\left.\mu\, \frac{d\la}{d\mu}\right|_{n\to 4} \, = \,
\be_\la^{(1)} \, = \, \frac{3\la^2}{(4\pi)^2}\,,
\qquad
\qquad
\la(\mu_0)=\la_0.
\label{RGfor_la}
\eeq
To first order in $\lambda$, the solution of (\ref{RGfor_la}) is
given by $\,\la (\mu)=\la_0 + \be_\la^{(1)}\,\ln
\left(\mu/\mu_0\right)$. The obtained result is just the physical
running coupling $\lambda=\lambda(\mu)$ after  $\mu\to |Q|$.
Inserting it in (\ref{effpo1}) and disregarding the ${\cal
O}(\la^3)$ terms, one obtains a similar expression for $V$ in which
$\mu$ is traded for ${\mu_0}$. In other words, this confirms that
$V_{scal}$ is RG-invariant to the order under consideration, as we
expected. On the other hand, the physically relevant
$\phi$-dependence is not affected by our manipulations, also as
expected. Similar considerations can be done for the massive case.

What is the lesson from this simple example? It is the following: in
order to gather some physical insight from he $\mu$-dependence, we
should not use Eqs. (\ref{effpo1}) and (\ref{RGfor_la}) at the same
time, but separately. No physical information can be extracted from
the cancelation of $\mu$-dependence in any sector of the EA, except
to confirm the consistency of the overall RG procedure. In contrast,
if one keeps the $\mu$-dependencies separately, then, in the first
case, this dependence is useful to derive the form of effective
potential for the massless field, whereas in the second case it
corresponds to the momentum-dependence in the scattering amplitudes
at high energies.

\vskip 2mm
%%%%%%%%%%%%%%%%%%%%%%%%%%%%%%%%%%%%%%%%%%%%%%%%%%%%%%%%%%%%%
%%%%%%%%%%%%%%%%%%%%%%%%%%%%%%%%%%%%%%%%%%%%%%%%%%%%%%%%%%%%%
%\vskip 1mm

{\Large $\bullet$} An essential point in the discussion of the
cosmological vacuum energy is the presence of an external metric
(the FLRW one in the standard cosmological scenario).
This presence is tremendously crucial for the RG applied to CC,
however it was not taken into account in the considerations
presented in~\cite{AG}.

In flat space, the one-loop vacuum contributions are given by just
closed loops of matter fields without external legs. It corresponds
to the $V_{vac}$ piece in Eq.\,(\ref{Veff}), and at one loop simply
reads (\ref{rfvacenergy2}). At any order, such loop contributions do
not depend on any physical quantity apart from the masses and
couplings of the matter particles. So, the $\mu$-dependence,
although mathematically present, has no physical implication. The
situation changes dramatically in curved space-time, where the
relevant Feynman diagrams include matter field loops with a number
of external legs of the metric\,\cite{PoImpo}. In this situation,
there are physically reasonable ``identifications'' of
$\mu$\,\cite{cosm}-\cite{Gruni}. \vskip 1mm

{\Large $\bullet$}  For example, in the approach used
in\,\cite{cosm,nova}, the MS-based RG was employed for all
parameters, including CC, $\la$ and $m$. Here $\mu$ was identified
with the $H$-dependent sharp cutoff $\mu\sim\rho_c^{1/4}\sim
\sqrt{H\,M_P}$. Therefore, the contributions from particles with
masses $m>\sqrt{H\,M_P}$ were excluded from the running of $\la$,
$m$ and CC. As we have discussed, the choice of the cutoff is not
unique because the status of the renormalization technique in this
context is not perfect at present, but the use of the RG was
consistent and technically correct. Similarly, we arrived at
(\ref{runphysCC}) through the alternative ansatz $\mu=H$, and this
is also technically correct within the corresponding subtraction
scheme (in fact, one which is conceptually more related to momentum
subtraction). See also \cite{Reuter1,Reuter2} for similar
identifications.

{\Large $\bullet$} Although the physical vacuum energy is
independent of $\mu$, the reason for choosing (``identifying'') a
particular value is because we don't know the full quantum structure
of the vacuum energy in curved space-time. Therefore, one can
estimate it by picking a representative energy parameter of the
system (in this case $H$, the expansion rate of the Universe). The
procedure is similar to computing a QCD cross-section at a given
order, say $\sigma(e^+e^-\to q\bar{q})$ at order ${\cal
O}(\alpha_{em}^2\alpha_s^2)$ in the electromagnetic and strong
coupling constants $\alpha_i=g_i^2/4\pi$. The result is explicitly
depending on $\mu$ since, at this order, one must renormalize the
strong coupling constant $\alpha_s(\mu^2)$. The $\mu$-dependence
would disappear, of course, if we could compute the cross-section to
all orders. However, to get a numerical estimate at a finite order,
the value of $\mu$ is typically ``identified'' with the
energy/momentum of the process ($\mu\to |Q|$), as this avoids the
effect from large logs. In the cosmological case, the situation is
more complicated; in part because the explicit computation is not
feasible in an intrinsically curved space-time background with
$\CC\neq 0$ (not even at the one-loop level), and in part also
because choosing $\mu$ is more subtle when we are working in the
infrared regime.

{\Large $\bullet$} Last, but not least, in curved space-time the
minimally interacting scalar field is not renormalizable (see, e.g.,
\cite{book}) and hence the MS-based RG can not be applied as it is
done in \cite{AG}. In the presence of the non-minimal
scalar-curvature term $R\phi^2$, the SSB becomes quite nontrivial
and the renormalization of the theory is more complicated
\cite{sponta}.
%%%%%%%%%%%%%%%%%%%%%%%%%%%%%%%%%%%%%%%%%%%%%%%%%%%%%%%%%%%%%
%%%%%%%%%%%%%%%%%%%%%%%%%%%%%%%%%%%%%%%%%%%%%%%%%%%%%%%%%%%%%
\vskip 1mm

In summary, a necessary condition for the physical running of the CC
is to have a non-trivial gravitational background.  However, we do
not know if such condition is sufficient. Hence, in order to
establish (or disprove) such running one has to derive explicitly
the quantum corrections, or at least indicate the form that they can
have. Since we cannot perform the explicit calculations for the time
being, in the next section we address only the possible form of the
quantum effects.

%% 5
%%%%%%%%%%%%%%%%%%%%%%%%%%%%%%%%%%%%%%%%%%%%%%%%%%%%%%%%%
\section{\large Possible forms of the quantum effects on the
vacuum action in curved space-time}

The possible form of the quantum corrections to the Einstein-Hilbert
action and to the CC term should be represented by some functional
of the metric and its derivatives, while the other fields are in the
stable vacuum states. Therefore, for the present consideration we do
not need to distinguish between the vacuum and induced parts of the
quantum corrections. Within QFT in curved space, the classical
action of vacuum has the form (\ref{total}) without the matter term,
where
\beq S_{HD} &=& \int d^4x \sqrt{-g}
\left\{a_1C^2+a_2E+a_3{\Box}R+a_4R^2 \right\}\,. \label{HD} \eeq
Here $\,C^2=R_{\mu\nu\al\be}^2 - 2 R_{\al\be}^2 + (1/3)\,R^2\,$ is
the square of the Weyl tensor and $\,E = R_{\mu\nu\al\be}^2 - 4
R_{\al\be}^2 + R^2\,$ is the integrand of the Gauss-Bonnet
topological term.

What could be the form of the quantum corrections to (\ref{HD})? The
first option is to perform explicit calculations of the quantum
corrections, including the finite part. The history of such
calculations goes back to the early works on quantum theory in
curved space \cite{stze71}. The most complete result in this
direction has been obtained in \cite{apco} through the derivation of
the Feynman diagrams in the framework of linearized gravity and also
by using the heat kernel solution of \cite{bavi90}. The output of
such calculation includes only the terms of zero, first and second
order in the curvature tensor. For example, the one-loop result for
the massive real scalar field reads
\beq {\bar \Ga}^{(1)}_{vac} &=& \frac{1}{2(4\pi)^2}\,\int d^4x
\,\sqrt{-g}\, \Big\{\,\frac{m^4}{2}\cdot\Big(\frac{1}{\ep}
+\frac32\Big) \,+ \,\Big(\xi-\frac16\Big)\,m^2R\,
\Big(\frac{1}{\ep}+1\Big) \nonumber
\\
&+& \frac12\,C_{\mu\nu\al\be} \,\Big[\frac{1}{60\,\ep}+k_W(a)\Big]
C^{\mu\nu\al\be} \,+\,R
\,\Big[\,\frac{1}{2\ep}\,\Big(\xi-\frac16\Big)^2\, +
k_R(a)\,\Big]\,R\,\Big\}\,.
\label{final}
\eeq
Here
${1}/{\ep}\equiv {2}/{(4-n)} +\ln({4\pi \mu^2}/{m^2}) - \ga_E$
is a parameter related to dimensional regularization, and $\xi$ is
the coefficient of the non-minimal coupling. It is clear that the
first term on the \textit{r.h.s.} of (\ref{final}) coincides with
the flat-space expression in Eq.\,(\ref{Vacfree2}). The full result
(\ref{final}), therefore, provides the
corresponding generalization for the case when there is a
non-trivial background. The non-local form factors have the form
\beq k_W(a) &=& \frac{8A}{15\,a^4}
\,+\,\frac{2}{45\,a^2}\,+\,\frac{1}{150}\,, \nonumber
\\
k_R(a) &=&
A{\tilde \xi}^2
+ \Big(\frac{2A}{3a^2}-\frac{A}{6}+ \frac{1}{18}\Big)
{\tilde \xi}
+ \frac{A}{9a^4}
- \frac{A}{18a^2}
+ \frac{A}{144} +
+ \frac{1}{108\,a^2} -\frac{7}{2160}
\,,
\label{W}
\eeq
where we used notations
\beq {\tilde \xi} = \xi-\frac16\,, \qquad
A\,=\,1-\frac{1}{a}\ln\,\Big(\frac{2+a}{2-a}\Big)\,, \qquad a^2 =
\frac{4\Box}{\Box - 4m^2}\,. \label{Aa} \eeq Let us note that
similar expressions have been obtained for massive fermion and
vector cases \cite{apco} and also for the scalar field background,
where the form factors were calculated for the $\,\phi^2 R\,$ and
$\,\phi^4\,$ terms \cite{bexi}. Looking at these expressions, it is
clear that it is no longer possible to use an effective potential
approach as in the flat space case considered in section 4, since
some of the new terms involve non-local contributions and are
explicitly dependent on the external momenta, including in the
gravitational sector. In the UV limit the nonlocal form factors have
the logarithmic behavior similar to (\ref{QED}), while in the IR
limit they follow the Appelquist and Carazzone theorem \cite{AC}.

The form factors (\ref{W}) contain important information about
quantum corrections. However, the CC and Einstein-Hilbert sectors of
(\ref{final}) carry no relevant form factors. In both cases there
are divergences, and the $\mu$-dependence is identical to the one of
the MS-based RG analysis that we have presented in section 4.
However, there is nothing real behind this formal
$\mu$-dependence\,\cite{Brown}. Contrary to the higher derivative
sectors, we do not observe physical running in the CC and G sectors
of the vacuum action. It is noticeable that, in these two cases,
there is no obvious correspondence between the RG approaches in the
MS and physical subtraction schemes, not only in the in IR, but also
in UV.
%%%%%%%%%%%%%%%%%%%%%%%%%%%%%%%%%%%%%%%%%%%%%%%%%%%%%%%%%%%%%%%%
The origin for this lack of correspondence is that, when the CC is
present, the space-time background is unavoidably curved; hence, the
expansion around the flat background,
$g_{\mu\nu}=\eta_{\mu\nu}+h_{\mu\nu}$, although computationally
feasible, it does miss part of the most relevant quantum
corrections. The passing to the flat space does not affect the form
factors of the higher derivative terms, but the ones for the CC and
Einstein-Hilbert terms just vanish, because $\Box\La=0$ and $\Box R$
is an irrelevant total derivative term. Hopefully, the problem can
be solved by performing calculations on some dynamical or at least
de Sitter background, but even in such relatively simple case it has
not been done yet. Such calculations look quite nontrivial, because
one needs to compute not only the divergences, but also the first
finite non-local correction using some (yet unknown) physical
renormalization scheme.

Coming back to the discussion on the viability of the CC running, we
can prove a ``no-go'' theorem concerning the possible/impossible
form of the relevant quantum corrections to the CC term. We can show
that these quantum effects cannot just be given by non-local terms,
if they are still polynomial in the curvatures. In order to support
this statement, let us consider some polynomial term in curvatures
having a finite number of Green functions insertions. These
insertions should correspond to the massive quantum field propagator
in curved space, but since the polynomial term admits the flat space
expansion, we can take the ${\cal O}(h_{\mu\nu})$ term for each
curvature and flat-space propagator. The simplest possible terms of
this sort are
\beq R_{\mu\nu}\,\frac{1}{\Box + m^2}\,R^{\mu\nu} \,\,,  \qquad
R\,\frac{1}{\Box + m^2}\,R\,, \label{nonloc} \eeq
and the other possible terms differ from these ones by a finite
number of curvature terms and corresponding Green functions.
Inasmuch as we are always interested in the large mass limit, we can
expand the propagator as
\beq \frac{1}{\Box + m^2} \sim \frac{1}{m^2}\, \Big(1 -
\frac{\Box}{m^2} + \,...\Big)\,. \label{propa} \eeq
The first term in the parenthesis just gives a contribution to the
local terms in the vacuum action (\ref{total}), whereas the second
term generates a ${\cal O}(H^6/m^4)$ contribution and is
phenomenologically irrelevant (notice that $\Box\sim H^2$ and $R\sim
H^2$ in the FLRW metric). Furthermore, the higher order curvature
terms give even smaller contributions. The upshot is that we have no
relevant running of CC and $1/G$ from the polynomial non-local
terms.

Does it mean that the running is impossible? No. Let us remember
that the relevant Feynman diagrams renormalizing the CC term consist
of closed loops of matter particles with an unrestricted number of
external gravitational legs and a corresponding infinite number of
Green function insertions of the matter fields. In other words, one
actually generates a series of infinite non-local products of
curvatures. The actual renormalization effect on the CC should come
from the resummation of this series, and the result is expected to
depend on a massless Green function, say  $G_0\sim 1/\Box$. For
instance, let us take the form $R\, {\cal F}(G_0\, R)$, for some
unknown function ${\cal F}$ of dimension $2$. The canonical
possibility would be ${\cal F}= M^2 G_0 R$, where $M$ is an
effective mass. It is easy to see that, in this case, the resummed
expression would be of order $M^2\,H^2$. Therefore, if  $M$ is not
far away from $M_P$, it would lead to the kind of (potentially
measurable) running law (\ref{runphysCC}) which we already expected
from general considerations of covariance\,\cite{nova}.

The positive sign that this kind of resummation is feasible is that,
in the unique case when we can find an exact solution for the EA in
curved space-time, that is to say, for the anomaly-induced action
(\ref{quantum for massive}), we do observe a very similar sort of
resummation into a massless propagator. The anomaly-induced action
(see, e.g., \cite{PoImpo} and references therein) has no sign of the
Green functions of the original quantum fields, but it leads to the
fourth order conformal operator $\Delta_4 = \nabla^4 +
2\,R^{\mu\nu}\nabla_\mu\nabla_\nu - \frac23\,R{\nabla^2} +
\frac13\,(\nabla^\mu R)\nabla_\mu$, which was not present in the
initial theory.

An indirect confirmation, through explicit calculation, that the
RG-like corrections to the CC and Einstein-Hilbert terms are
possible, can be obtained within the method suggested in
\cite{Shocom,asta} (see also \cite{fossil}). We refer to the
mentioned papers
%% (see also \cite{PSW})
for details and just
present the final result at the one-loop level:
\beq
{\bar \Gamma}
&=&
-\,\int d^4 x\sqrt{-{\bar g}} \,e^{2\si}\, [{\bar R}+6({\bar
\na}\si)^2] \,\cdot\, \Big[\, \frac{1}{16\pi G} - f\cdot\si\,\Big]
\nonumber
\\
\nonumber
\\
&-& \int d^4 x\sqrt{-{\bar g}}\,e^{4\si}\,\cdot\,
\Big[\frac{\La}{8\pi G}\,-\,g\cdot\si\,\Big] \,+\, \mbox{\sl higher
derivative terms.} \label{quantum for massive} \eeq The EA
(\ref{quantum for massive}) is written using a special conformal
parametrization of the metric $g_{\mu\nu}=e^{2\si(x)}{\bar
g}_{\mu\nu}$, similar to Eq.\,(\ref{conf_rep}). Here $g$ and $f$ are
the MS-based $\be$-functions for the corresponding parameters of the
higher derivative action (\ref{HD}). The above expression is a
generalization of the RG corrected classical vacuum action
(\ref{total}), in a perfect correspondence with our considerations
in section 3.

Let us stress that the result (\ref{quantum for massive}) has been
obtained without the polynomial expansion in curvatures and is
essentially based on the non-covariant separation of the conformal
factor of the metric. As can be seen, in this parametrization the
``running'' of CC and $1/G$ with $\sigma$ is perfectly possible. In
this sense, this result represents a strong argument in favor of the
possibility of the relevant quantum corrections in the CC and
Einstein-Hilbert sectors. The reason is that the expression
(\ref{quantum for massive}) does not depend on the flat-expansion
approach and that is why it enables one to reproduces the RG-based
result.

%% 6
%%%%%%%%%%%%%%%%%%%%%%%%%%%%%%%%%%%%%%%%%%%%%%%%%%%%%%%%%%%%
\section{\large Conclusions}

\qquad We have discussed some important theoretical aspects of the
CC or vacuum energy density, most particularly the possibility that
it can be a running quantity in QFT. In view of the magnificent
observational status of cosmology at present, we believe that such
discussion is timely and highly convenient to encourage the
theoretical community to focus more deeply on the QFT status of the
CC term rather than replacing it too often with ad hoc scalar fields
and other artificial variables. While the technical limitations of
the current QFT methods can not prove that the CC running takes
place, they intriguingly suggest that it is a most natural scenario
and that it should be further investigated with renewed efforts. At
the same time, these methods cannot disprove such running either.
Therefore, all positive statements here must be done with proper
caution; and the same is true, of course, for the negative
statements. In particular, trying to prove a ``no running'' theorem
on the basis of $\mu$-independence of EA \cite{AG} is misleading,
because the $\mu$-independence is nothing else but the universal
property of EA, which can not support any constructive statement
about the physical running of the CC. The latter, if really exists,
should manifest as a functional dependence on the cosmological
variables of the expanding background (such as the Hubble rate,
scale factor etc). The role of $\mu$ is merely formal and has no
direct physical meaning at all. However, the fact that $\mu$
parameterizes the structure of the quantum corrections (as is
well-known e.g. in QED and QCD) can be the clue to use the RG also
in cosmology as the optimal instrument to search for the possible
form of the quantum corrections to the vacuum energy and extract
their concrete dependence on the physical quantities of the
expanding Universe. In our opinion, there is no doubt that this is
one of the main challenges for future investigations on the
fundamental theoretical aspects of this discipline, and it can be
the clue to a most sought-after interface, if not the very
touchstone, that may link Particle Physics and
Cosmology\,\cite{nova}.

Finally, in the present situation, wherein the perspective to
obtain a solid QFT result remains still unsettled, it is
absolutely legitimate to use the phenomenological approach to
investigate the implications of a running CC and to derive the
cosmological restrictions on it (see e.g. the extensive work
along these lines presented in references \cite{Reuter2} and
\cite{cosm}-\cite{fossil}).

%\vspace{0.1cm}
\section*{Acknowledgments}
The authors are thankful to E. Gorbar, B. Guberina, R. Horvat, H.
\v{S}tefan\v{c}i\'{c} and E. Verdaguer for helpful discussions. The
work of IS has been supported by CNPq, FAPEMIG, FAPESP and ICTP. The
work of JS has been supported in part by MEC and FEDER under project
FPA2007-66665, by  DIUE/CUR Generalitat de Catalunya under project
2009SGR502 and  by the Spanish Consolider-Ingenio 2010 program CPAN
CSD2007-00042; JS also thanks the MPI in Munich for the hospitality
and financial support while part of this work was being carried out,
and also to FAPEMIG for the support provided during a visit at the
Fed. Univ. of Juiz de Fora.

%\newpage
%%%%%%%%%%%%%%%%%%%%%%%%%%%%%%%%%%%%%%%%%%%%%%%%%%%%%%

\end{document}